\documentclass{aa}
\usepackage{graphicx}
\usepackage{txfonts}
\usepackage{natbib}
\def   \aj {{\rm {AJ}}}
\def   \araa {{\rm {ARA\&A}}}
\def   \apj {{\rm {ApJ}}}

\def   \apjs {{\rm {ApJS}}}
\def   \apss {{\rm {Ap\&SS}}}
\def   \aap {{\rm {A\&A}}}
\def   \aapr {{\rm {A\&AR}}}

\def   \mnras {{\rm {MNRAS}}}

\def   \apjl{\rm {ApJL}}
\def   \nat{\rm {Nat.}}

\begin{document} 
   \title{Inhomogeneous cloud coverage through the Coulomb explosion of dust in substellar atmospheres}
   \author{C. R. Stark
          \inst{1},
          Ch. Helling \inst{1},
          \and
          D. A. Diver\inst{2}\fnmsep
          }
   \institute{SUPA, School of Physics and Astronomy, University of St Andrews, St Andrews, KY16 9SS, Scotland, UK.\\
              \email{craig.stark@st-andrews.ac.uk} \and SUPA, School of Physics and Astronomy, Kelvin Building, University of Glasgow, Glasgow, G12 8QQ, Scotland, UK.}
 \date{}
   \abstract
   {Recent observations of brown dwarf spectroscopic variability in the infrared infer the presence of patchy cloud cover.}
   {This paper proposes a mechanism for producing inhomogeneous cloud coverage due to the depletion of cloud particles through the Coulomb explosion of dust in atmospheric plasma regions. Charged dust grains Coulomb-explode when the electrostatic stress of the grain exceeds its mechanical tensile stress, which results in grains below a critical radius $a<a^{\rm Coul}_{\rm crit}$ being broken up.}
   {This work outlines the criteria required for the Coulomb explosion of dust clouds in substellar atmospheres, the effect on the dust particle size distribution function, and the resulting radiative properties of the atmospheric regions.}
   {Our results show that for an atmospheric plasma region with an electron temperature of $T_{e}=10$~eV ($\approx10^{5}$~K), the critical grain radius varies from $10^{-7}$ to $10^{-4}$~cm, depending on the grains' tensile strength. Higher critical radii up to $10^{-3}$~cm are attainable for higher electron temperatures. We find that the process produces a bimodal particle size distribution composed of stable nanoscale seed particles and dust particles with $a\geq a^{\rm Coul}_{\rm crit}$, with the intervening particle sizes defining a region devoid of dust.  As a result, the dust population is depleted, and the clouds become optically thin in the wavelength range $0.1-10~\mu$m, with a characteristic peak that shifts to higher wavelengths as more sub-micrometer particles are destroyed.}
   {In an atmosphere populated with a distribution of plasma volumes, this will yield regions of contrasting radiative properties, thereby giving a source of inhomogeneous cloud coverage. The results presented here may also be relevant for dust in supernova remnants and protoplanetary disks.}
    \titlerunning{electrostatic disruption in substellar atmospheres}
 \maketitle
%
\section{Introduction}
A defining characteristic of brown dwarfs is that their atmospheres are cool enough that dust particles can condense and form large-scale cloud structures~\citep{helling2004,dehn2007,helling2006,helling2008b,helling2008c,witte2009,witte2011,tsuji2002,allard2001,burrows1997,marley2002,morley2012}.  Cloud formation depletes the ambient gas and alters the objects' resultant electromagnetic spectrum (e.g. see recent review by~\citet{helling2014}).  Observationally, the presence of clouds has been associated with photometric variability, which is believed to be caused by inhomogeneous cloud coverage~\citep{bailer2001,artigua2009,radigan2012,gillon2013,girardin2013,buenzli2014}.  However, such variability could also be connected to magnetic activity~\citep{littlefair2008,metchev2013}, such as asymmetrically distributed, magnetically induced spots~\citep{scholz2005}.  Recent observations of Luhman 16B's infrared spectroscopic variability have allowed a two-dimensional map of the brown dwarf's surface to be constructed.  It reveals large-scale bright and dark features supporting the idea of patchy cloud cover~\citep{crossfield2014}.  This cloud variability can be explained using a spot model where thick and thin cloud regions at different temperatures (indicating different atmospheric heights) provide the required inhomogeneity~\citep{radigan2012,apai2013,burgasser2014,morley2014}.  To quantify the inhomogeneity, recent observations have probed the horizontal and vertical structure of the atmosphere, identifying pressure-dependent phase variations that underlie the observed variability~\citep{buenzli2012,biller2013}.    

Inhomomgeneous cloud coverage may be significant with respect to the transition from L-type to T-type objects (e.g. see~\cite{burgasser2014}).  Ultracool L and T dwarfs bridge the divide between cool stars and giant gas planets, and understanding the transition from L-type dusty atmospheres to dust-free T-type atmospheres is a continuing problem (e.g.~\cite{tinney2003,faherty2012,dupuy2012,burgasser2014}). In the latter case, the atmosphere may not be cloud free, but rather the clouds lie below the level where the atmosphere becomes optically thick.  Differing models have been invoked to explain the transition, including cloud depletion due to changes in atmospheric chemistry~(e.g. \cite{knapp2004,tsuji2005,burrows2006}), the formation of holes in the cloud layers~(e.g. \cite{ackerman2001,burgasser2002,gelino2002,allard2013,crossfield2014,morley2014}), and due to atmospheric circulation~\citep{freytag2010,showman2013}.  Furthermore, a growing body of theoretical and observational evidence~(e.g.~\cite{hallinan2008,route2012,mclean2012,osten2009,antonova2013,berger2009}) suggests that ionization processes can be significant in these dusty environments producing extensive regions of plasma~\citep{helling2011a,helling2013,bailey2013,stark2013}.  This implies that substellar atmospheres could be composed of dusty gas-plasma mixtures, underlining the importance of atmospheric dusty plasma processes and their role in cloud evolution.    

A plasma is defined as a quasineutral gas of charged and neutral particles which exhibit collective behaviour~\citep{chen1984}.  Locally, the entirety of the gas does not need to be ionized for it to behave as a plasma.  As long as the collective, long-range electromagnetic influence of the many distant charged particles dominates over short-range, binary interactions the gas only needs to be weakly ionized (i.e. partially ionized, $f_{e}\gtrsim10^{-7}$) for it to exhibit plasma behaviour~\citep{diver2013,fridman2008}.  In substellar atmospheres, Alfv\'{e}n ionization can produce localized volumes of plasma with degrees of ionization, $f_{e}$, ranging from $10^{-6}$ to as large as $1$, depending on the atomic and molecular species that is ionized~\citep{stark2013}.  Alfv\'{e}n ionization occurs when a stream of neutral gas (that reaches a critical threshold speed) impinges on a low-density magnetized plasma. The inflowing neutrals scatter the plasma ions, leaving behind a significant charge imbalance of electrons that are accelerated by their collective self-repulsion to ionising energies~\citep{alfven1960,diver2005}.  In principle, the volumes of atmospheric plasma created can range from small to large scales, potentially populating a significant fraction of the atmosphere.

\cite{helling2011b} investigated the effect of dust-induced collisional ionization in substellar atmospheres using \textsc{Drift-Phoenix} model atmosphere results.  In this process, if the energy imparted from one dust grain to another during a collision exceeds the grain's work function, an electron will be liberated. Turbulence-induced dust-dust collisions were found to be the most efficient dust-dust collisional ionization process that could enhance the local degree of ionization, but on its own it was insufficient to form a plasma.  The degree of ionization can be further enhanced by atmospheric electrical discharge events~\citep{helling2011a,helling2013,bailey2013,craig2009,craig2013}:  an ensemble of inter-grain discharges, in analogy to laboratory microdischarges~\citep{becker2006}, or large-scale cloud lightning will greatly increase the electron number density.  Furthermore, the bombardment of the atmosphere by cosmic rays~\citep{rimmer2013} will also contribute to the formation of atmospheric plasmas.  

The properties of the produced plasma, such as the electron and ion temperature,  will depend on the ionization process which created it.  Due to their significantly lower mass relative to the neutrals and ions, the plasma electrons only lose a small fraction of their energy following a collision.  As a result, the electrons and the heavier species will not necessarily be in thermal equilibrium and can have differing temperatures.  In typical terrestrial discharges $T_{e}>T_{\rm gas}\approx T_{i}$, where $T_{i}$ is the ion temperature and electron temperatures are $T_{e}\approx1-100$~eV ($\approx10^{4}-10^{6}$~K)~\citep{fridman2008,diver2013}.  In substellar atmospheres electrons from thermal ionization are in thermal equilibrium with the ambient neutral gas ($T_{e}\approx T_{\rm gas}$) and have electron temperatures of $T_{e}\approx \mathcal{O}(0.01-0.1$~eV$)$ ($\approx \mathcal{O}(10^{2}-10^{3}$~K$)$). 

Contemporary atmospheric models of brown dwarf and exoplanetary (substellar) atmospheres do not exploit collective plasma processes (such as dust growth via plasma deposition) and how these effects fit into a global understanding of cloud formation and other atmospheric phenomena (such as large-scale flows).  A major challenge is to model cloud formation and its consequent effect on the surrounding environment, involving plasma processes such as non-spherical dust growth~\citep{stark2006} and electrical discharge events~\citep{craig2013}.  \cite{opik1956} investigated the electrostatic disruption (or Coulomb explosion) of dust grains, where the electrostatic stress of a body holding a net charge exceeds its mechanical tensile strength (the maximum stress or pressure that a material can withstand while being pulled), resulting in it fracturing.  See also~\citet{rhee1976},~\citet{hill81},~\citet{fom1992},~\citet{ros1992},~\citet{gron2009}, among others.  

In this paper we outline a model to contribute to the understanding of inhomogeneous cloud coverage in substellar atmospheres via the Coulomb explosion of cloud dust particles.  Consider an atmosphere with plasma regions populated with dust cloud particles.  The spatial extent and distribution of the plasma regions will depend upon the ionization process that produced them and environmental factors.  For example, inhibited plasma transport perpendicular to the ambient magnetic field can result in anisotropically shaped plasma regions, structured by the field geometry.  In a dusty atmosphere, cloud particles may grow in the plasma environments themselves, grow in the region prior to plasma generation, or have migrated into the region via gravitational settling, wind flows or another transport mechanism.  As the dust charges, the accumulated electrostatic stress on the particles may overcome their mechanical tensile strength resulting in them breaking up and being destroyed.  Electrostatic disruption preferentially affects smaller grain sizes and this may have an effect on the particle size distribution function affecting the thickness and the electromagnetic properties of the cloud.  Depending on the distribution of the dusty plasma regions on the substellar surface, the electrostatic disruption of dust clouds could yield regions of contrasting cloud properties, thereby giving a source of inhomogeneous cloud coverage that could be associated with the observed spectroscopic variability.   
   
The aim of this paper is to investigate the conditions under which the electrostatic disruption of cloud dust particles occur in brown dwarf or gas giant planet atmospheres and to assess the contribution of Coulomb explosions to spectroscopic variability.  To do so we will use an example atmosphere from the \textsc{Drift-Phoenix} model atmosphere grid characterized by $\log{g}=3.0$, $T_{\rm eff}=1600$~K.  This paper is structured as follows: in Section~\ref{drift} we discuss contemporary substellar atmosphere and cloud formation models, specifically the \textsc{Drift-Phoenix} model; in Section~\ref{dust_plasma} the concept of dust in plasmas and grain charging is introduced.  Moreover, we consider the most appropriate plasma sheath model required to describe the charging of the dust and the region surrounding the dust grain;  Section~\ref{sec_coul} formulates the electrostatic disruption criteria of spherical dust grains in substellar atmospheres using the \textsc{Drift-Phoenix} model atmosphere discussed in Section~\ref{drift}.  It investigates the effect electrostatic disruption has on the dust particle size distribution, compares the timescale for disruption to dust growth timescales and discusses the observational consequences of Coulomb explosions;  Section~\ref{discussion} summarizes our findings and discusses the impact of the electrostatic disruption of dust on the subsequent chemistry that occurs in substellar atmospheres.  Note that unless otherwise stated, the equations in this paper are in SI units.

\section{Substellar model atmospheres \label{drift}}
  
The \textsc{Drift-Phoenix} model atmosphere code~\citep{helling2004,dehn2007,helling2006,helling2008b,helling2008c,witte2009,witte2011} is a 1D phase-non-equilibrium model, that describes the formation of dust and clouds in a self-consistently calculated atmosphere.  \textsc{Drift-Phoenix} model atmospheres incorporate radiative transfer, convective energy transport, chemical equilibrium, hydrostatic equilibrium~\citep{hauschildt1999} and dust cloud formation~\citep{woitke2003,woitke2004}. The cloud model numerically solves a system of differential equations that describe the time-dependent dust formation process, where seeds form (TiO$_{2}$) from a highly supersaturated gas.   Once the initial seed particles have formed, further growth proceeds via surface chemical reactions growing a grain mantle.  The \textsc{Drift-Phoenix} model atmospheres consider $32$ surface reactions and is restricted to seven participating solids:  TiO$_2$[s], Al$_2$O$_3$[s], Fe[s], SiO$_2$[s], MgO[s], MgSiO$_3$[s] and Mg$_2$SiO$_4$[s].  \textsc{Drift-Phoenix} solves the dust moment and element conservation equations, returning the parameterized particle size distribution function $f(a,z)$ of grain radius $a$ at atmospheric height $z$; and the chemical composition of the dust expressed in volume fractions $V_{s}/V_{\rm tot}$, for all solids $s$ involved where $V_{\rm tot}$ is the total dust volume.  

In this paper we are interested in the electrostatic disruption of dust grains in substellar atmospheres.   We consider a representative model atmosphere characterized by $\log{g}=3.0$, $T_{\rm eff}=1600$~K, applying results from the \textsc{Drift-Phoenix} model atmosphere simulations.  We are particularly interested in the chemical composition of the dust particles (hence the material volume fractions) and their size distribution since this informs of their mechanical properties and the particle sizes affected.  For this model atmosphere Fig.~\ref{fig_vs} shows the dust volume fraction $V_{s}/V_{\rm tot}$ and the mean grain size $\langle a\rangle$ as a function of atmospheric gas pressure $p_{\rm gas}$.  The nucleation-dominated upper cloud deck ($p_{\rm gas}\approx10^{-11}$~bar) predominantly contains small grains ($\langle a\rangle\approx10^{-7}$~cm) composed of TiO$_{2}$[s], whereas the lower atmosphere ($p_{\rm gas}\approx10^{-2}$~bar) is mainly made up of large dust grains ($\langle a\rangle\approx10^{-5}$~cm) composed of multiple species including Al$_{2}$O[s], Fe[s], Mg$_{2}$SiO$_{4}$[s] and trace amounts of other solid materials.  At mid-atmospheric pressures ($p_{\rm gas}\approx10^{-6}$~bar) the mean grain size is $\langle a\rangle\approx10^{-6}$~cm and the grain composition is more mixed being composed mainly of MgSiO$_{3}$[s], MgSiO$_{4}$[s], SiO$_{2}$[s], Fe[s] and MgO[s].  

\section{Dust cloud particles in substellar plasmas\label{dust_plasma}}
In this section we discuss the charging and the behaviour of dust immersed in a substellar atmospheric plasma region and consider the appropriate model required to best describe the plasma sheath of the dust.  

\subsection{Dust charging in plasmas\label{charge}}
Consider a dust particle submerged in an electron-ion plasma.  We will assume that the particle currents drawn from such a plasma are the only contributors to the grain charge and we ignore other charging processes such as photoionization and secondary electron emission.  This assumption is valid in the absence of significant UV radiation, which would liberate the electrons from the surface of the grain, possibly leading to positively charged grains~\citep{shukla2001}. Due to the greater mobility of the electrons relative to the ions, the dust becomes negatively charged initiating the formation of a plasma sheath (an electron depleted region) around the grain.  As the negative charge on the grain builds up, the number of electrons having the appropriate kinetic energy to overcome the grain potential, striking its surface, decreases.  Ions are subsequently accelerated from the bulk plasma, through the sheath and are ultimately deposited on the grain surface altering its shape, size, mass, charge and hence the potential of the grain.  As a result the number of electrons reaching the surface increases further altering the surface charge.  This occurs until the grain reaches an equilibrium state where the flux of electrons and ions at the grain surface is equal.  At this point the grain reaches the floating potential $\phi_{f}$ and the sheath length is of the order of the plasma Debye length $\lambda_{D}=(\epsilon_{0}k_{B}T_{e}/(n_{e}e^{2}))^{1/2}$.  The Debye length is the spatial length scale at which exposed charge is screened by the plasma. In this particle-flux equilibrium state we have a negatively charged grain that is screened by the bulk plasma such that on length scales greater than the Debye length, the electric field from the grain is zero.  The spatial extent of the sheath (the sheath length, $d_{\rm sh}$) occurs where the electric potential energy of the electrons is approximately equal to their thermal energy and the electric field from the grain is largely shielded from the rest of the bulk plasma.  .

For conducting grains the captured charge will distribute itself on the surface in the lowest energy configuration such that the electric field inside the conductor is zero, otherwise charges see a net force and move seeking equilibrium.  For spherical dust grains this leads to the uniform distribution of electric charge on the grain surface and the electric field can be simply described.  However, in ultra-cool substellar atmospheres dust grains are insulating and so once the electrons strike the surface their transport is inhibited and they can not easily reach the lowest energy configuration.  However, in a plasma the dust grains will be constantly bombarded by an isotropic flux of electrons and ions which allows the lowest energy configuration to be obtained even though electron transport in the dust grain is inhibited.  For spherical dust, the electric field from the grain will be the same as that from an equivalent conducting grain allowing the same formulae to be used to describe its electric field.

\subsection{Collisions in plasma sheaths\label{sec_coll}}
The effect of collisions in the plasma sheath defines how the sheath will behave and how it should be described.  In this section the significance of collisions in the sheath is quantified.  The plasma sheath surrounding a dust particle is considered collisionless if the collisional length scale (the mean free path, $\lambda_{\rm mfp,n}$) for ions with the neutral species is greater than the sheath extent, $d_{\rm sh}$: $\delta=d_{\rm sh}/\lambda_{\rm mfp,n}<1$; otherwise, the sheath is collisional~\citep{lieberman2005}.  In other words, if the ions do not participate in neutral collisions while transiting the sheath then $\delta<1$; if the ions experience multiple collisions then $\delta>1$.  In the case $\delta<1$ the effect of short-range binary collisions is negligible in comparison to long-range collective phenomena and the sheath is well described by solutions to the Vlasov equation~\citep{bouchoule1999}.  If the collisional scale length for ions is less than the sheath length then a collisional model describing the sheath is preferable.
Assuming that the sheath length is approximately  equal to the Debye length, $d_{\rm sh}\approx\lambda_{D}$ and $\lambda_{\rm mfp,n}=(n_{\rm n}\sigma_{\rm n})^{-1}$, we can write, 
\begin{equation}
\delta_{\rm ni}\approx\left(\frac{\epsilon_{0}k_{B}T_{e}}{n_{e}e^{2}}\right)^{1/2}n_{n}\sigma_{n}, \label{collision}
\end{equation}
where $n_{e}$ is the plasma electron number density, $n_{n}$ is the neutral species number density and $\sigma_{n}$ is the corresponding collisional cross-sectional area.  Note that in \textsc{Drift-Phoenix}, $n_{\rm gas}$ refers to the number density of all species and $n_{n}$ refers to the number density of the neutral species only, such that $n_{\rm gas}=n_{n}+n_{e}+n_{i}=n_{n}+2n_{e}$.  Thus, the degree of ionization takes the following form, $f_{e}=n_{e}/(n_{n}+n_{e})=n_{e}/(n_{\rm gas}-n_{e})$.  For ion-neutral interactions, we assume $\sigma_{n}\approx \pi r_{\rm n}^{2}$ where $r_{\rm n}\approx10^{-8}$~cm is the radius of a neutral species; therefore, 
\begin{equation}
\delta_{\rm ni}\approx10^{-15}\frac{\left(1-f_{e}\right)}{f_{e}^{1/2}\left(1+f_{e}\right)}\left(n_{\rm gas}T_{e}\right)^{1/2},\label{delta_ni}
\end{equation}
where $n_{\rm gas}$ is in units of [cm$^{-3}$].  The ratio of the sheath length and the ion-neutral mean free path, $\delta_{\rm ni}$, (Equation~\ref{delta_ni}) is plotted in Figure~\ref{delta_1} as a function of atmospheric gas pressure, $p_{\rm gas}$, for an example substellar atmosphere characterized by $\log{g}=3$ and $T_{\rm eff}=1600$~K (see Section~\ref{drift}).  Note that the atmospheric gas pressure $p_{\rm gas}$, in this plot and in subsequent plots, acts as a proxy for atmospheric height and in general does not reflect the true local atmospheric gas pressure. This is a result of varying the degree of ionization from the values calculated in \textsc{Drift-Phoenix} due to thermal ionization, to reflect additional non-thermal ionization processes that occur. As a consequence the local atmospheric gas pressure would change and deviate from its original \textsc{Drift-Phoenix} calculated value. Further to this, note that for $f_{e}=1$, $\delta_{\rm ni}=0$ since $n_{n}\rightarrow0$ and $\lambda_{\rm mfp,n}\rightarrow\infty$.  For low degrees of ionization ($f_{e}\approx10^{-7}$) the neutrals will be the majority species and the transport of the plasma ions will be inhibited by frequent collisions with the neutrals as they travel through the partially ionized medium.  If the degree of ionization increases and more of the gas is ionized, the neutral number density decreases and the likelihood of an ion-neutral interaction is smaller relative to the previous case.  In the case of a fully ionized region of plasma ($f_{e}=1$) there will be no neutrals for the ions to interact with and the sheath will be collisionless.  For atmospheric gas pressures $p_{\rm gas}\approx10^{-12}-10^{-4}$~bar in the model considered, the collisionless sheath criterion is satisfied, $\delta_{\rm ni}<1$, for all cases of $f_{e}$ and $T_{e}$ that characterize the plasma regions in the atmosphere.  At higher atmospheric gas pressures ($p_{\rm gas}\gtrsim10^{-3}$~bar) only plasmas with degrees of ionization $f_{e}\gtrsim10^{-4}$ satisfy $\delta_{\rm ni}<1$. In plasma regions where $\delta_{\rm ni}>1$, the sheaths are more accurately described by collisional models; however, $\delta_{\rm ni}>1$ is only obtained for $p_{\rm gas}\gtrsim10^{-3}$~bar, which is a very small fraction of the atmosphere; hence, we are justified assuming a collisionless sheath. 

\subsection{OML and Bohm theory\label{sec_oml}}
Following Section~\ref{sec_coll}, if we assume a collisionless sheath and a dust particle of radius $a$, the parameter $\xi=a/d_{\rm sh}$ defines the plasma sheath regime and the appropriate theory required to describe it: Bohm or Orbital Motion Limited theory (OML)~\citep{allen1957}.  Bohm theory ($\xi\gg1$) assumes that all ions that enter the sheath reach the dust surface and contribute to its charge, whereas OML theory ($\xi\ll1$) takes into account that some ions may be deflected and fail to reach the particle surface.  Assuming for the sheath length that  $d_{\rm sh}\approx\lambda_{D}$ we have,
\begin{eqnarray}
\xi&\approx& a\left(\frac{n_{e}e^{2}}{\epsilon_{0}k_{B}T_{e}}\right)^{1/2} \\ 
&\approx&0.1~a \left(\frac{f_{e}}{1+f_{e}}\right)^{1/2} \left(\frac{n_{\rm gas}}{T_{e}}\right)^{1/2}, \label{xi}
\end{eqnarray} 
where in the final expression $n_{\rm gas}$ is in units of [cm$^{-3}$] and $a$ is in units of [cm].  For $\xi\gg1$ the sheath extent is much smaller relative to the particle size and so the surface area of the dust particle is effectively the same as the sheath surface area.  In this regime, all ions that are accelerated from the bulk plasma and through the sheath reach the particle's surface contributing to its charge and potential.  For $\xi\gg1$ the sheath is called a Bohm sheath and the floating potential is given by~\citep{bouchoule1999},
\begin{equation}
\phi_{f}=-\frac{k_{B}T_{e}}{2e}\ln{\left(\frac{m_{i}}{2\pi m_{e}}\right)}.\label{bohm}
\end{equation}
Calculating the floating potential using Bohm theory gives an upper limit to the actual potential value.  In the regime $\xi\ll1$ the sheath length is much greater than the particle size.  In this scenario, it is more likely that ions entering the sheath with a random velocity component may not collide with the particle surface since its surface area is much smaller than the sheath's.  Instead, ions will have non-zero angular momentum and may participate in stable orbits around the dust particle or be ejected from the sheath on parabolic orbits.  In this case the floating potential, $\phi_{f}$ is reduced and is calculated using OML theory and is given by the expression~\citep{bouchoule1999},
 \begin{equation}
\exp{\left(\frac{e\phi_{f}}{k_{B}T_{e}}\right)}=\left(\frac{T_{i}m_{e}}{T_{e}m_{i}}\right)\left(1-\frac{e\phi_{f}}{k_{B}T_{i}}\right). \label{oml}
\end{equation}
It is important to note that the floating potential in both cases depends on the bulk properties of the plasma.  The parameter $\xi=a/d_{\rm sh}$ (Equation~\ref{xi}) is plotted in Figure~\ref{xi_1}.  Predominantly $\xi<1$ indicating that OML theory is best suited to describing plasma sheaths in substellar atmospheres.  Since the neutral number density increases with atmospheric gas pressure $p_{\rm gas}$, the electron number density also increases with $p_{\rm gas}$ for a given degree of ionization.  As a result, the Debye length decreases with increasing electron number density and $\xi$ increases in value as the atmospheric pressure increases.  Holding the electron number density fixed, increasing (decreasing) the electron temperature has the effect of increasing (decreasing) the Debye length and decreasing (increasing) $\xi$.

There is no universal consensus on the dynamical theory of ions in plasma sheaths and the resulting effect on dust charging in plasmas; therefore, we present results from both Bohm and OML theory in order to give a pragmatic representation.  Bohm theory gives a valuable upper limit to the floating potential whereas OML represents an indication of how the non-zero angular momentum of ions in the sheath affect the charging of the dust particle. For the results presented here, the difference in the floating potentials calculated using the Bohm and OML is never more than a factor $\approx2$.

\section{Inhomogenous cloud coverage through the Coulomb explosion of dust clouds \label{sec_coul}}

This section calculates the criteria for charged dust grains to Coulomb explode and the effect this has on the resultant cloud particle size distribution function in substellar atmospheres. 
 
\subsection{Criteria for electrostatic disruption of spherical dust grains\label{disrupt}}

Dust immersed in a plasma becomes negatively charged through the collection of plasma electrons.  During the charging process described in Section~\ref{dust_plasma} the dust can be electrostatically disrupted if the resultant electrostatic stress acting on the dust grain (due to the captured electrons) overcomes the mechanical tensile strength (the maximum stress that a material can withstand while being pulled) of the particle, breaking it apart.  Consider a conducting spherical dust grain of radius $a$ with a uniform tensile strength $\Sigma$ (measured in Pascals, Pa).  Electrostatic disruption occurs when the electrostatic stress ($\epsilon_{0}E^{2}/2$) on the grain exceeds its tensile strength~\citep{opik1956},
\begin{equation}
\frac{\epsilon_{0}E^{2}}{2}=\frac{\epsilon_{0}}{2}\left(\frac{Q}{4\pi\epsilon_{0}a^{2}}\right)^{2}\geq\Sigma,
\end{equation}
therefore, using $Q=Ne$, 
\begin{equation}
Ne\geq\pi(32\epsilon_{0}\Sigma)^{1/2}a^{2}\label{n_eqn}.
\end{equation}
This expression states that there is a maximum number of net charges that can reside on the surface of a spherical dust grain of radius $a$ without it breaking up.  Equivalently, this expression can be stated in terms of the surface charge density $\sigma\geq(2\epsilon_{0}\Sigma)^{1/2}$, which is independent of the grain size.  For the critical charge density $\sigma_{c}$, grains with a smaller surface area can only tolerate a smaller number of charges in comparison to larger grains.  
Therefore, in a plasma where spherical dust particles are charged to the floating potential $\phi_{f}$ ($=Ea$), the result can be recast in terms of the critical cloud particle size, $a_{\rm crit}^{\rm Coul}$, below which the dust particles are electrostatically disrupted,
\begin{equation}
\frac{\epsilon_{0}}{2}\left(\frac{\phi_{f}}{a}\right)^{2}\geq\Sigma,
\end{equation}
hence, 
\begin{equation}
a\leq a_{\rm crit}^{\rm Coul}=\left(\frac{\epsilon_{0}}{2\Sigma}\right)^{1/2}\phi_{f} \label{acrit}
\end{equation}
Therefore, dust particles that are charged and have a size below $a<a_{\rm crit}^{\rm izedCoul}$ will Coulomb explode and be broken up.  For multi-composite materials (as expected for cloud particles in substellar atmospheres) the Rule of Mixtures (RoM) can be used to approximate the resultant tensile strength $\Sigma$ of a hybrid material:
\begin{equation}
\Sigma=m_{b}\sum_{s=1}^{n}(V_{s}/V_{\rm tot})\Sigma_{s}, \label{upper}
\end{equation}
where $V_{s}/V_{\rm tot}$ and $\Sigma_{s}$ are the volume fraction and tensile strength of the $s^{\rm th}$ material of the composite~\citep{callister1985,torquato1999,hsieh2005,kim2001,li1999,torquato2000,raabe1995}.  The RoM can be used to approximate properties of composite materials such as the elastic modulus.  Additional physical effects can be incorporated into the RoM formulation such as interfacial bonding between the composite materials; and the effect of particle size distributions and particle clustering within the composite~\citep{li2001}.  To quantify the effect of imperfect bonding between the individual components and the departure from perfectly formed idealized crystalline solids the parameter $m_{b}\in(0,1]$ is introduced.

\begin{figure}
\resizebox{\hsize}{!}{\includegraphics{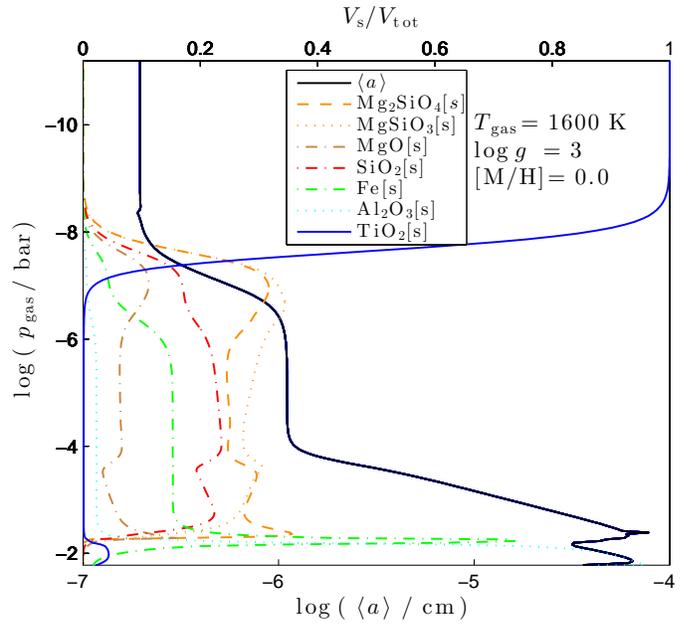}}
\caption{Volume fraction $V_{s}/V_{\rm tot}$ and the mean grain size $\langle a\rangle$ of a mineral cloud in a substellar atmosphere with $T_{\rm eff}=1600$~K and $\log{g}=3$.  Results are from the \textsc{Drift-Phoenix} model atmosphere simulation and is consistent with \cite{helling2008c}. \label{fig_vs}}
\end{figure}

\begin{figure}
\resizebox{\hsize}{!}{\includegraphics{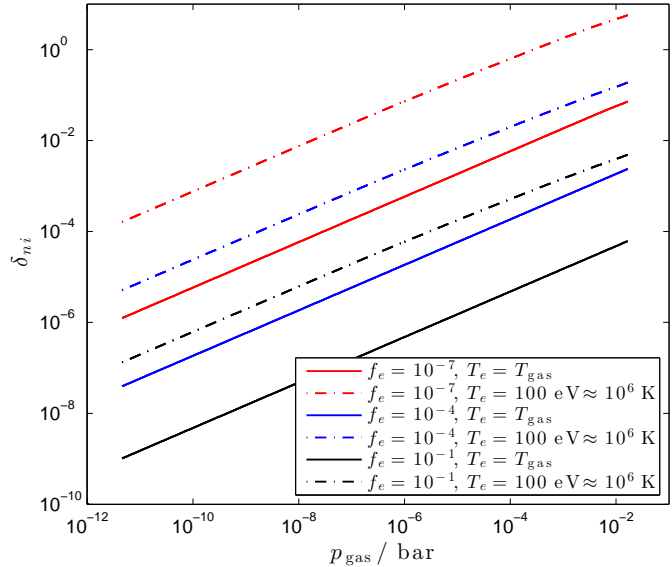}}
\caption{Ratio of the sheath extent to the ion-neutral mean free path $\delta_{\rm ni}=d_{\rm sh}/\lambda_{\rm mfp,n}$ as a function of atmospheric gas pressure $p_{\rm gas}$ for the same model atmosphere as in Fig.~\ref{fig_vs}.  $\delta_{\rm ni}$ is plotted for degrees of ionization $f_{e}=10^{-7}$, $10^{-4}$ and $10^{-1}$ and electron temperatures $T_{e}=T_{\rm gas}$ ($\approx  \mathcal{O}(10^{2}-10^{3}$~K$)$) and $T_{e}=100$~eV ($\approx10^{6}$~K), indicating the extremities of $\delta_{\rm ni}$ in parameter space.\label{delta_1}}
\end{figure}

\begin{figure}
\resizebox{\hsize}{!}{\includegraphics{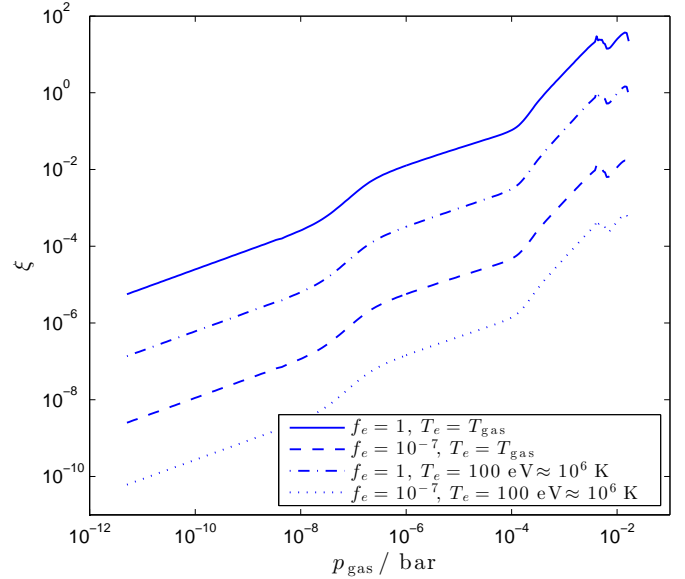}}
\caption{Ratio of the dust grain radius to the sheath extent $\xi=a/d_{\rm sh}$ as a function of atmospheric gas pressure $p_{\rm gas}$ for the same model atmosphere as in Fig.~\ref{fig_vs}.  $\xi$ is plotted for $a=\langle a\rangle$, for degrees of ionization $f_{e}=10^{-7}$ and $1$ and electron temperatures $T_{e}=T_{\rm gas}$ ($\approx  \mathcal{O}(10^{2}-10^{3}$~K$)$) and $T_{e}=100$~eV ($\approx10^{6}$~K) representing the bounds of parameter space.  \label{xi_1}}
\end{figure}

\begin{figure}
\resizebox{\hsize}{!}{\includegraphics{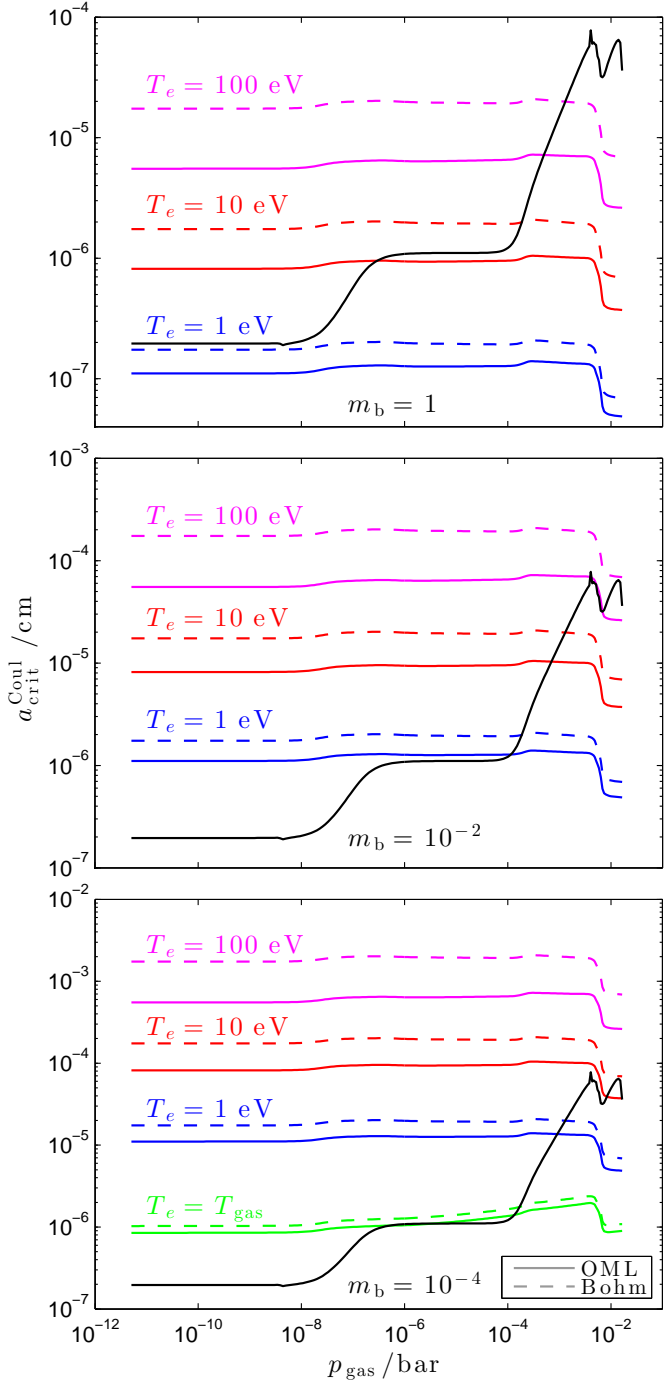}}
\caption{$a_{\rm crit}^{\rm Coul}$ as a function of atmospheric gas pressure $p_{\rm gas}$ for the same substellar atmosphere as in Fig.~\ref{fig_vs}.  $a_{\rm crit}^{\rm Coul}$ is plotted for electron temperatures $T_{e}=T_{\rm gas}$ ($\approx  \mathcal{O}(10^{2}-10^{3}$~K$)$, green), $1$~eV ($\approx10^{4}$, blue), $10$~eV ($\approx10^{5}$, red) and $100$~eV ($\approx10^{6}$, magenta).  The plot also shows $a_{\rm crit}^{\rm Coul}$ calculated using Bohm (dash) and OML (solid) theory to obtain the floating potential $\phi_{f}$.   The mean grain size $\langle a\rangle$ as a function of gas pressure $p_{\rm gas}$ for the example substellar atmosphere considered here is over plotted (black).\label{a_plots_1}}
\end{figure}

Using the values of tensile strengths listed in Table~\ref{tbl-2} and the \textsc{Drift-Phoenix} volume fractions $V_{s}$ (Figure~\ref{fig_vs}), Figure~\ref{a_plots_1} shows the critical cloud particle size, $a_{\rm crit}^{\rm Coul}$, calculated using Equation~\ref{acrit} and using the tensile strength, $\Sigma$, from Equation~\ref{upper}.  The critical cloud particulate size is plotted as a function of local gas pressure $p_{\rm gas}$ for electron temperatures ranging from $1$ to $100$~eV.  The top, middle and bottom subplots show $a^{\rm Coul}_{\rm crit}$ for $m_{\rm b}=1,~10^{-2}$ and $10^{-4}$ respectively. For $m_{\rm b}=10^{-4}$, electron temperatures $T_{e}=T_{\rm gas}$ ($\approx  \mathcal{O}(10^{2}-10^{3}$~K$)$, green) are considered but are not considered for the cases $m_{\rm b}=1$ and $10^{-2}$, since the critical particle sizes are too small and are unphysical.

The critical particle size is dependent upon the floating potential and the tensile strength of the dust cloud particle, $a_{\rm crit}^{\rm Coul}\propto\phi_{f}/\sqrt{\Sigma}$.  For both sheath models (Bohm and OML) the floating potential depends upon the electron temperature, $T_{e}$.  As $T_{e}$ increases the plasma electrons become more energetic and so a greater number of electrons can reach the surface of the grain, overcoming the electrostatic potential and increasing the ultimate potential reached when the particle-flux equilibrium configuration is achieved.  For the Bohm model the floating potential (and hence $a_{\rm crit}^{\rm Coul}$) has a simple linear dependence upon $T_{e}$ and so an increase in $T_{e}$ is reflected in $\phi_{f}$ and $a_{\rm crit}^{\rm Coul}$  

For example when $m_{\rm b}=1$, at $p_{\rm gas}\approx10^{-4}$~bar the critical particle radius varies from $a_{\rm crit}^{\rm Coul}\approx10^{-7}$~cm when $T_{e}=1$~eV ($\approx10^{4}$~K) to $a^{\rm Coul}_{\rm crit}\approx10^{-6}$~cm when $T_{e}=10$~eV ($\approx10^{5}$~K).  Similarly the critical radius increases to $a_{\rm crit}^{\rm Coul}\approx10^{-5}$~cm when $T_{e}=100$~eV ($\approx10^{6}$~K).  In comparison to the Bohm model, the floating potential and hence $a_{\rm crit}^{\rm Coul}$ calculated with OML theory is smaller for a given $T_{e}$; however,  the difference in $a_{\rm crit}^{\rm Coul}$ for the two models is never more than a factor $\approx2$.  

For a given electron temperature, the variation of $a_{\rm crit}^{\rm Coul}$ with $p_{\rm gas}$ is predominantly determined by the variation in the total tensile strength of the dust particles, $\Sigma$, hence the volume fraction $V_{s}/V_{\rm tot}$ and individual tensile strengths of the constituent materials.  As an example to elucidate this point, we look at the values of $a_{\rm crit}^{\rm Coul}$ for $m_{\rm b}=1$ and the $T_{e}=10$~eV ($\approx10^{5}$~K), Bohm theory solution (red-dash) in Figure~\ref{a_plots_1}.  At the bottom of the atmosphere ($p_{\rm gas}\approx10^{-2}$~bar), Figure~\ref{fig_vs} shows that Al$_{2}$O$_{3}$[s] is the dominant dust component ($V_{s}/V_{\rm tot}\approx1$) which has a relatively high tensile strength ($241.4$~MPa) and this raises the overall mechanical tolerance of the dust.  As a result, the critical particle radius ($\propto\Sigma^{-1/2}$) has a relative low value, $a_{\rm crit}^{\rm Coul}\approx10^{-7}$~cm.  However, as we climb through the cloud ($p_{\rm gas}\approx10^{-4}$~bar) the volume fraction of Al$_{2}$O$_{3}$[s] decreases drastically and other materials, such as MgSiO$_{3}$[s], that have a significantly lower tensile strength ($9.6$~MPa) dominate the mechanical properties of the grains, therefore lowering the electrostatic stress that can be mechanically tolerated.  As a consequence $a_{\rm crit}^{\rm Coul}$ increases in value from that at higher atmospheric pressures to $a_{\rm crit}^{\rm Coul}\approx10^{-6}$~cm.  Finally, at the top of the cloud the grains are predominately composed of TiO$_{2}$[s], which has a tensile strength similar to magnesium silicates ($36.41$~MPa) and so the critical particle size is similar in value $a_{\rm crit}^{\rm Coul}\approx10^{-6}$~cm. 

The material tensile strengths used (Table~\ref{tbl-2}) are obtained from material science, where they are obtained under optimum laboratory conditions using idealized samples and present an upper bound to their actual value.  In reality the tensile strengths of the dust particles will be much lower reflecting the imperfect bonding and structure of the particles.  Although the dust particles could be crystalline in nature (see~\cite{reitmeijer2009}) the possibility exists that they are not and are only a loose aggregate of weakly bonded material. The tensile strength is the macroscopic quantity that represents the effect of the microscopic bonds holding the atoms of the solid together. An estimation of the tensile strength of a dust grain can be made by considering the work done, $W$, by the bonds to maintain a stable solid structure: $\Sigma\approx W/V_{\rm d}$, where $V_{\rm d}=4\pi a^{3}/3$ is the volume of the dust grain.  If a solid is held together by $N_{\rm b}$ bonds of energy $E_{\rm b}$, then $W\approx E_{\rm b}N_{\rm b}$. The number of bonds can be estimated by calculating the number of participating atoms of radius $r_{0}$ in a cross-sectional slice through the dust grain of area $\pi a^{2}$; hence, $N_{\rm b}\approx(a/r_{0})^{2}$, where $r_{0}\approx10^{-8}$~cm is the classical radius of an atom.  Therefore, the tensile strength of the dust grain can be estimated as $\Sigma\approx (a/r_{0})^{2}E_{\rm b}/(4\pi a^{3}/3)$. For a crystalline spherical dust grain of radius $a\approx10^{-4}$~cm and typical lattice energy $E_{\rm b}\approx 10$~eV, the estimated tensile strength is $\Sigma\approx10$~MPa, which is consistent with the tensile strengths listed in Table~\ref{tbl-2}. However, for a similarly sized dust grain that is weakly bonded, held together by van der Waals bonds ($E_{\rm b}\approx0.01$~eV), where only 1\% of the atoms participate in bonding, the tensile strength can be as low as $\Sigma\approx10^{-4}$~MPa. Therefore, in comparison to the typical tensile strengths in Table~\ref{tbl-2}, $\Sigma\approx\mathcal{O}(1-100$~MPa), the parameter $m_{\rm b}$ can be as low as $10^{-4}$.

The effect of imperfect bonding is demonstrated in Figure~\ref{a_plots_1}, for values of $m_{b}=10^{-2}$ (middle subplot) and $10^{-4}$ (bottom subplot).  Consider $a_{\rm crit}^{\rm Coul}$ calculated for $T_{e}=10$~eV ($\approx10^{5}$~K) using Bohm theory, at $p_{\rm gas}\approx10^{-4}$~bar, assuming a perfectly bonded crystalline dust particle ($m_{b}=1$), $a_{\rm crit}^{\rm Coul}\approx2\times10^{-6}$~cm.  For $m_{b}=10^{-2}$ this value of the critical dust radius increases by a factor $m_{b}^{-1/2}\approx10$ to become $a_{\rm crit}^{\rm Coul}\approx2\times10^{-5}$~cm.  Similarly, for $m_{b}=10^{-4}$, $a_{\rm crit}^{\rm Coul}$ increases by a factor of $\approx100$  such that $a_{\rm crit}^{\rm Coul}\approx2\times10^{-4}$~cm. For a $T_{e}=100$~eV plasma the critical particle radius can be as large as $a^{\rm Coul}_{\rm crit}\approx10^{-3}$~cm when $m_{\rm b}=10^{-4}$.

Charged dust grains with a radius below the critical radius $a<a_{\rm crit}^{\rm Coul}$, will be disrupted and broken up, leaving a population of dust grains with $a>a_{\rm crit}^{\rm Coul}$.  In addition, there will be a further population of crystalline seed particles composed of a single substance (e.g. TiO$_{2}$, Fe, Al$_{2}$O$_{3}$, etc) with a stable radius, $a_{\rm seed}=a_{\rm crit}^{\rm Coul}=(\epsilon_{0}/(2\Sigma_{s}))^{1/2}\phi_{f}$, where $m_{b}=1$. For example, pure crystalline seed particles composed of TiO$_{2}$ ($\Sigma=36.21$~MPa) immersed in a plasma with $T_{e}=1$~eV ($\approx10^{4}$~K) will have a minimum critical radius $a_{\rm seed}\approx10^{-7}$~cm.  Consider an aggregate of crystalline seed particles forming a spherical particle cluster.  In this context the parameter $m_{\rm b}$ is no longer unity and contains a number of additional effects, such as the packing efficiency of arranging constituent particles into a desired volume. For simple cubic, body-centred cubic and face-centred cubic stacking geometries the packing fraction is $f_{\rm p}\approx$ 0.52, 0.68 and 0.74, respectively.  The packing fraction sets an upper limit to the value of $m_{\rm b}$. Assuming that the aggregate's tensile strength reflects that of the crystalline seed particles, the ratio of the tensile strengths for the two cases is $1/m_{\rm b}$ and the ratio of the critical radius is $1/\sqrt{m_{\rm b}}$.  Therefore, if the particle aggregate's radius is smaller than $a_{\rm seed}/\sqrt{m_{\rm b}}$, the aggregate is unstable to electrostatic disruption and the unstable range is defined by $] a_{\rm seed},a_{\rm seed}/\sqrt{m_{\rm b}}]$.  However, the tensile strength of the aggregate may be lower than that of the constituent seed particles since they may be bonded together via van der Waals interactions.   For example, consider a particle aggregate of radius $10^{-5}$~cm composed of $10^{-7}$~cm seed particles bonded together via van der Waals interactions ($\approx0.01$~eV), the tensile strength of the aggregate can be $10^{-3}$~MPa. In this scenario the ratio of tensile strengths will be smaller, resulting in a greater critical radius for stability and a larger region of instability.

 The seed particles that compose the particle aggregate are initially charged to the floating potential $\phi_{f}=Q_{\rm seed}/(4\pi\epsilon_{0}a_{\rm seed})$, where $Q_{\rm seed}$ is the charge on a seed particle.  The resulting particle aggregate will also be charged to the floating potential such that $\phi_{f}=Q_{\rm agg}/(4\pi\epsilon_{0}a_{\rm agg})$, where $Q_{\rm agg}$ and $a_{\rm agg}$ is the charge and radius of the particle aggregate respectively. Therefore, the charge ratio is $Q_{\rm agg}/Q_{\rm seed}=a_{\rm agg}/a_{\rm seed}$. The total charge of the aggregate can also be calculated by summing the charge from the individual seed particles. The number of seed particles in the aggregate, $N_{\rm seed}$, is given by the volume occupied by the seed particles in the aggregate, $V_{\rm agg}f_{\rm p}$, divided by the volume of a seed particle, $V_{\rm seed}$; therefore, $N_{\rm seed}=(a_{\rm agg}/a_{\rm seed})^{3}f_{\rm p}$. The total charge of the aggregate is $Q_{\rm tot}=(a_{\rm agg}/a_{\rm seed})^{3}f_{\rm p}Q_{\rm seed}$, which is greater than the aggregate charge calculated from the floating potential, $Q_{\rm tot}>Q_{\rm agg}$.  The process of particle aggregation results in charge being lost by the aggregate back into the plasma.
 
However, this analysis assumes only surface charges; if the particle aggregate has a volumetric charge  distribution, the electric field will be non-zero inside the aggregate and the constituent charged seed particles will see a non-zero electrostatic force.  This will undermine the mechanical stability of the aggregate and will reduce the value of $m_{\rm b}$ further. This is in contrast to a conducting dust grain, where the electric charge will distribute itself on the surface such that the electric field inside is zero and the electric potential is constant. The surface area of the aggregate, $4\pi a_{\rm agg}^{2}$, is less than the total available surface area, $N_{\rm seed}4\pi a_{\rm seed}$.  Assuming that each seed particle is identically charged, the charge on the surface of the aggregate is given by $Q_{\rm surf}=a_{\rm seed}Q_{\rm agg}/(a_{\rm agg}f_{\rm p})$ and the internal charge by $Q_{\rm int}=Q_{\rm agg}-Q_{\rm surf}$. For example, assuming $a_{\rm seed}=10^{-7}$~cm, $a_{\rm agg}=10^{-5}$~cm and $f_{\rm p}\approx0.5$, the surface charge is $Q_{\rm surf}\approx10^{-2}Q_{\rm agg}$.  Therefore, the majority of the electric charge is held within the body of the aggregate. Let's consider that the net electric charge is distributed uniformly throughout the volume of the aggregate so that the electric field as a function of radius, $r$, inside the aggregate is $E=Q_{\rm agg}r/(4\pi\epsilon_{0}a^{3}_{\rm agg})$. Comparing the surface electric field and the internal electric field of the aggregate yields $E/E_{\rm surf}=r/a_{\rm agg}$.  This allows a comparison of the relative strength of the internal electric field.  Note that in the case of a conducting dust grain the internal electric field is zero. On length scales comparable to the seed particle size,  $E/E_{\rm surf}=a_{\rm seed}/a_{\rm agg}\approx10^{-2}$ for $a_{\rm seed}=10^{-7}$~cm and $a_{\rm agg}=10^{-5}$~cm.  Therefore, a porous dust aggregate will have a repulsive, non-zero electrostatic energy density that weakens its mechanical integrity.  The analysis of dust grains collected from the Jupiter-family comet 67P/Churyumov-Gerasimenko show the dust particles are fluffy, porous aggregates with very low tensile strengths of the order of $10^{3}$~Pa ~\citep{schulz2015}.  This supports the tensile strength values considered in this paper and that the dust grains may be a loose, porous, fluffy aggregate of weakly bonded material.

Therefore, in a dusty plasma region there exists an instability strip for particle sizes between seed particles and particle aggregates composed of multiple seed particles, defined by $] a_{\rm seed},a_{\rm seed}/\sqrt{m_{\rm b}}]$.  In order for stable grains with a radius greater than $a_{\rm seed}/\sqrt{m_{\rm b}}$ to form they cannot do so through aggregation nor thin film surface growth via neutral gas-phase surface deposition.  The accretion of unstable fragments that produce a bigger stable particle could occur as long as the accretion timescale is shorter than the timescale for the fragment to Coulomb explode.  If the plasma is only present in certain regions of the atmosphere, dust growth can occur classically and can be transported into the plasma region, whereupon the unstable particles will Coulomb explode upon charging to reach the floating potential.

Electron field emission occurs when a dust grain's surface electric field becomes large enough that electrons are liberated from its surface. If significant, the resulting current must be taken into account to determine the grain's net charge from the steady-state balance of currents to its surface.  For dust grains, electron field emission becomes a significant process when a critical electric field strength at the grain's surface is reached $E_{\rm fe}=\phi/a\gtrsim10^{7}$~V~cm$^{-1}$~\citep{gomer1961,mendis1974,draine1979,ishihara2007,mann2014}, where $\phi$ is the grain potential and $a$ is the radius of the dust grain.  This value is calculated from the Fowler-Nordheim equation, which determines the electron current drawn from a conducting dust grain with a work function, $W_{\rm f}$ (typically $W_{\rm f}=1-5$~eV).  Assuming $\phi/a=10^{7}$~V~cm$^{-1}$, field emission is not a significant grain charging process if $a\gtrsim10^{-7}\phi$.  In a $T_{e}=1$~eV plasma, $\phi_{f}=\mathcal{O}(1$~V) and so dust grains with $a\gtrsim10^{-7}$~cm are not significantly affected by field emission, likewise $a\gtrsim10^{-6}$~cm and $a\gtrsim10^{-5}$~cm for plasmas with $T_{e}=10$~eV ($\phi_{f}=\mathcal{O}(10$~V)) and $100$~eV ($\phi_{f}=\mathcal{O}(100$~V)) respectively. Only for the smallest dust grains in a $T_{e}=100$~eV plasma is field emission a significant contributor to dust charging and so inhibit the attainment of $a^{\rm Coul}_{\rm crit}$ for electrostatic disruption.  For this grain population, the potential they acquire is restricted to $\phi=10^{7}a$; therefore, for $a=10^{-6}$~cm ($10^{-7}$~cm), the potential reached is $\phi=10$~V ($\phi=1$~V). As a result, the electrostatic disruption criterion for the grains revert to that for the case $\phi_{f}\approx\mathcal{O}(10$~V) ($\phi_{f}\approx\mathcal{O}(1$~V)); hence, when $T_{e}\approx\mathcal{O}(10$~eV) ($T_{e}\approx\mathcal{O}(1$~eV)).  Therefore, the grain population affected by field emission may still be able to participate in electrostatic disruption.  For example, when $m_{\rm b}=10^{-4}$ and taking $\phi_{f}=\mathcal{O}(10$~V) ($T_{e}=10$~eV) as the grain potential reached, $a_{\rm crit}^{\rm Coul}=10^{-4}$~cm and so dust grains with $a=10^{-6}$~cm (i.e. those affected by field emission when $T_{e}=100$~eV) will still fracture.  Furthermore, in substellar atmospheres dust grains are primarily composed of insulating material and so the work function can be larger e.g. $W_{\rm f}=10$~eV~\citep{miloch2009}, yielding larger tolerable values of $E_{\rm ef}$ before field emission becomes a significant process. 

\subsection{Timescale considerations\label{timescales}}

The electrostatic disruption of dust in substellar plasmas will only have a significant effect on the particle size distribution if the timescale for disruption dominates over other dust growth and destruction timescales.  In a fully ionized plasma where there are no neutrals, nucleation and growth as described by contemporary substellar atmosphere and cloud formation models is not suitable since the chemistry is predicated upon neutral gas-phase chemistry.  In this case an extended approach of dust nucleation and growth in substellar plasmas is required.  For a fully ionized plasma region, the timescale for electrostatic disruption precludes the timescale for other processes associated with charged dust and so electrostatic disruption will occur.  In a partially ionized plasma, the ionized regions will be composed of gas-plasma mixtures and so if the gas is not sufficiently ionized then the nucleation and growth rate for dust formation could be greater than the rate of destruction via disruption.  

\begin{figure}
\resizebox{\hsize}{!}{\includegraphics{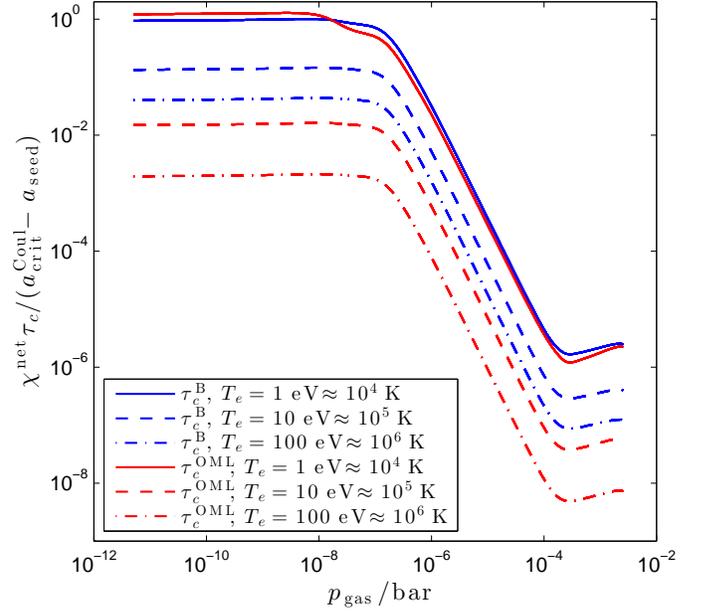}}
\caption{Comparison of the competing timescales of dust growth and dust charging, evaluated for $f_{\rm e}=10^{-7}$, $m_{\rm b}=1$ and a range of electron temperatures.\label{chi_t_2}}
\end{figure}

The characteristic charging time for dust in a plasma is given by~\citep{bouchoule1999},
\begin{eqnarray}
\tau_{c}^{\rm B}&=&A^{-1}, \label{t_bohm} \\
\tau_{c}^{\rm OML}&=&\frac{1}{A}\left[\frac{T_{e}}{T_{i}}+\left(\frac{T_{e}}{T_{i}}\right)^{1/2}(1-y_{0})\right]^{-1}, \label{t_oml} \\
A&=&\frac{an_{e}e^{2}}{4\epsilon_{0}(m_{i}k_{B}T_{e})^{1/2}},
\end{eqnarray}
where $\tau^{\rm B}_{c}$ is the Bohm theory characteristic charging time for dust; $\tau_{c}^{\rm OML}$ is the OML theory charging time; and $y_{0}=-e\phi_{f}/(k_{B}T_{e})$. Once seed dust particles are formed in substellar atmospheres they grow as they gravitationally settle until they ultimately evaporate at the base of the atmosphere.
The grain-size dependent, equilibrium settling speed is obtained from the balance between the gravitational force and the frictional force exerted by the surrounding atmospheric gas on the dust grain~\citep{woitke2003}.  The net growth velocity for heterogeneous growth is $\chi^{\rm net}$ [cm~s$^{-1}$] (positive for growth; negative for evaporation) which includes all the growth velocities for each participating condensate, $\chi^{\rm net}=\sum_{s}\chi^{s}_{\rm net}$~\citep{helling2008c}.  In \textsc{Drift-Phoenix} the growth process occurs through the physical adsorption of impinging gaseous molecules on the total grain surface. The molecules diffuse across the grain surface via hopping to a solid island of suitable composition where surface chemistry and chemisorption occurs, creating a new unit of material in the solid's crystal structure~\citep{helling2006}. Note that the growth process is independent of the grain charge.  Consider a seed particle in an atmospheric region that has just become partially ionized.  If the seed particle can grow fast enough it may be able to surmount the critical radius $a_{\rm crit}^{\rm Coul}$ and escape electrostatic disruption.  As a result the particle size distribution function will not be truncated significantly by the process of electrostatic disruption.  For a net growth velocity $\chi^{\rm net}$, the time required for a particle to grow to a size $a_{\rm crit}^{\rm Coul}$ is given by
\begin{equation}
\tau_{\rm crit}^{\rm gr}=\frac{a_{\rm crit}^{\rm Coul}-a_{\rm seed}}{\chi^{\rm net}}.
\end{equation}
Therefore, for electrostatic disruption to have a significant effect on the particle size distribution function we require the charging time to be less than the growth time ($\tau_{c}<\tau_{\rm crit}^{\rm gr}$),
\begin{equation}
\frac{\tau_{c}}{\tau^{\rm gr}_{\rm crit}}=\frac{\chi^{\rm net}\tau_{c}}{a_{\rm crit}^{\rm Coul}-a_{\rm seed}}< 1. \label{time}
\end{equation}
Figure~\ref{chi_t_2} shows Equation~\ref{time} evaluated as a function of atmospheric gas pressure using Equations~\ref{t_bohm} and~\ref{t_oml} for the dust charging time and $\chi^{\rm net}$ from \textsc{Drift-Phoenix}.  Figure~\ref{chi_t_2} presents the worst case scenario for dust charging, where the atmospheric gas is just sufficiently ionized (i.e. $f_{e}=10^{-7}$) to exhibit collective plasma effects, hence $\tau_{\rm c}$ is at its largest; and the dust grains are very mechanically strong (i.e. $m_{\rm b}=1$) resulting in $a^{\rm Coul}_{\rm crit}$ having the smallest possible values and the quickest dust growth timescales, $\tau^{\rm gr}_{\rm crit}$.  In general, for all cases and all atmospheric gas pressures $\tau_{\rm c}/\tau^{\rm gr}_{\rm crit}\lesssim 1$, so the seed dust grains charge quicker than the timescale for growth and the particle size distribution function will be significantly truncated by the electrostatic disruption of dust particles. As the atmospheric gas pressure increases, the gas-phase neutral number density increases and so for a given degree of ionization, the electron number density increases resulting in a smaller charging time and $\tau_{c}\ll \tau_{\rm crit}^{\rm gr}$. For higher degrees of ionization, $f_{e}\geq10^{-7}$, the charging time decreases, becoming much faster than the net growth speed of the dust grains and the effect of electrostatic disruption is significant in the evolution of the particle size distribution function.

\subsection{Effect on the dust particle size distribution function\label{dist}}

Cloud particles of size $a<a_{\rm crit}^{\rm Coul}$, immersed in a plasma, are subject to fracturing by electrostatic disruption. This can lead to a bimodal distribution of final grain sizes, comprising particles larger than $a_{\rm crit}^{\rm Coul}$ and smaller than nano-sized seed particles (both of which are stable to disruption), with intermediate sizes expunged.  This could yield regions of contrasting cloud properties in the atmosphere, thereby giving a source of inhomogeneous cloud coverage.  To understand the role of electrostatic disruption in substellar atmospheres the effect on the dust particle size distribution function needs to be quantified.

\begin{figure}
\resizebox{\hsize}{!}{\includegraphics{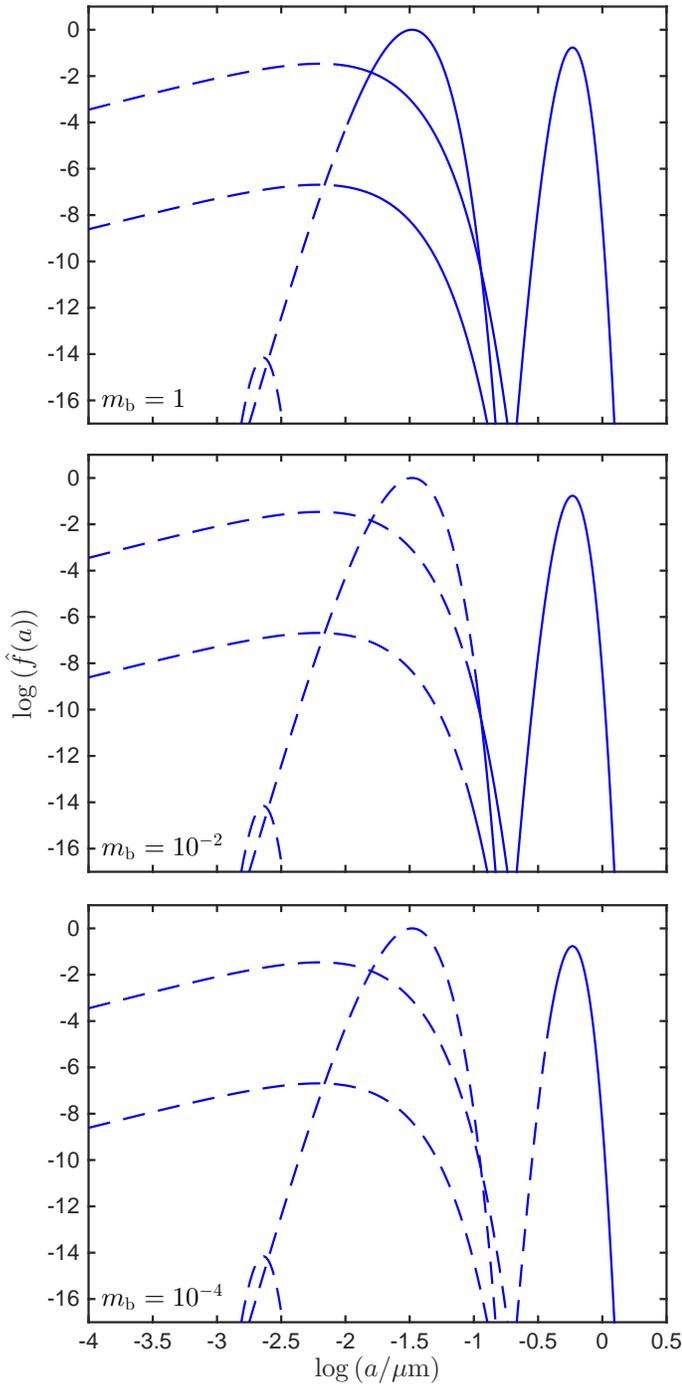}}
\caption{Normalized potential exponential particle size distribution function $\hat{f}(a)$ as a function of particle radius $a$ for various atmospheric heights. The distribution function is shown for atmospheric pressures $p_{\rm gas}\approx10^{-8}$,~$10^{-6}$,~$10^{-5}$~$10^{-4}$ and $10^{-2}$~bar, to exemplifying its evolution with atmospheric height.  Regions of the size distribution function that are Coulomb unstable are denoted with a dashed line, whereas regions of the size distribution that are Coulomb stable are denoted by a solid line. \label{dist_func_1}}
\end{figure}

\begin{figure}
\resizebox{\hsize}{!}{\includegraphics{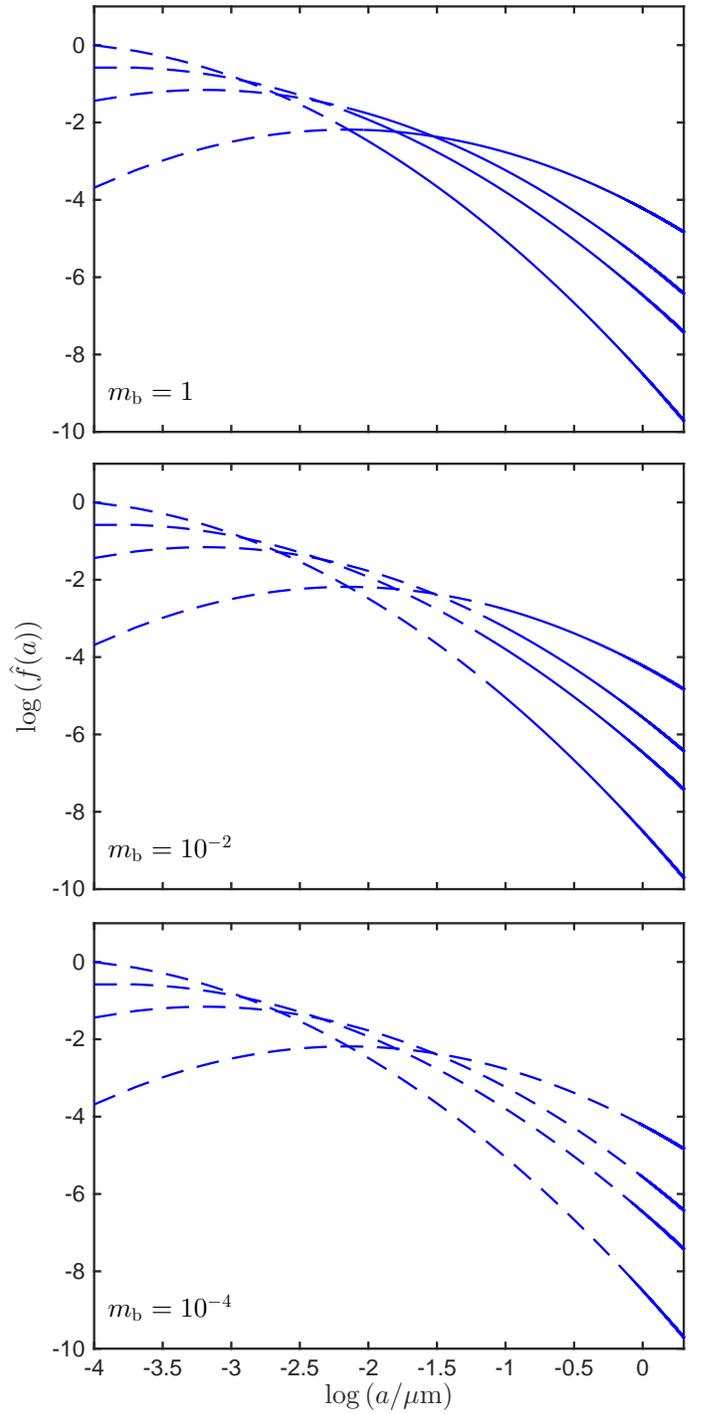}}
\caption{Normalized lognormal particle size distribution function $\hat{f}(a)$ as a function of particle radius $a$ for various atmospheric heights. The distribution function is shown for atmospheric pressures $p_{\rm gas}\approx10^{-9}$,~$10^{-6}$,~$10^{-4}$ and $10^{-2}$~bar respectively, to exemplifying its evolution with atmospheric height.  Regions of the size distribution function that are Coulomb unstable are denoted with a dashed line, whereas regions of the size distribution that are Coulomb stable are denoted by a solid line. \label{dist_func_logn_1}}
\end{figure}

Figures~\ref{dist_func_1} and~\ref{dist_func_logn_1} show how the particle size distribution function $f(a,z)$ as a function of cloud particle size $a$, varies through the atmosphere, where $z$ is the atmospheric height.  Seed dust particles form high in the atmosphere, the dust settles under the effect of gravity as it grows and then finally evaporates at the base of the atmosphere.  As a consequence, as the grains grow, the peak of the particle size distribution function migrates towards large particle sizes as the atmospheric gas pressure increases. Figures~\ref{dist_func_1}~and~\ref{dist_func_logn_1} shows two potential particle size distribution functions derived using the dust moments, $K_{j}=\int_{0}^{a_{\rm max}}a^{j}f(a,z)$d$a$, from \textsc{Drift-Phoenix}: a potential exponential size distribution~\citep{helling2008c} and a lognormal size distribution. The lognormal distribution is included to increase the impact and relevance of the results presented here to other models where such a distribution function is assumed. Using the \textsc{Drift-Phoenix} dust moments a lognormal particle size distribution function can be given by
\begin{equation}
f(a,z)=\frac{1}{a\sigma\sqrt{2\pi}}\exp{\left[-\frac{(\ln{(a)}-\mu)^{2}}{2\sigma^{2}}\right]},
\end{equation}
where $\mu$ and $\sigma$ are the mean and variance of the distribution respectively and are given by,
\begin{eqnarray}
\mu&=&\ln{\left[\frac{ \langle a\rangle^{2}}{\sqrt{\langle A\rangle+\langle a\rangle^{2}}}\right]}, \\
\sigma&=&\frac{1}{2}\ln{\left(\frac{\langle A\rangle}{\langle a\rangle^{2}}+1\right)}, 
\end{eqnarray}
where
\begin{eqnarray}
\langle a\rangle&=&K_{1}/K_{0}, \\
\langle A\rangle&=&4\pi K_{2}/K_{0},
\end{eqnarray}
and $\langle A\rangle$ is the mean dust surface area~[cm$^{2}$].  In Figure~\ref{dist_func_1} and~\ref{dist_func_logn_1} each curve represents the dust particle size distribution function at different atmospheric heights, starting at the cloud top where $p_{\rm gas}\approx10^{-11}$~bar and finishing at the bottom of the cloud where $p_{\rm gas}\approx10^{-3}$~bar.  The top, middle and bottom subplots show results for $a^{\rm Coul}_{\rm crit}$ when $m_{\rm b}=1$, $10^{-2}$ and $10^{-4}$ respectively, for the case $T_{e}=10$~eV (Bohm theory).  As $m_{b}$ decreases and $a^{\rm Coul}_{\rm crit}$ increases, larger regions of the particle size distribution $f(a,z)$ become Coulomb unstable and deeper cloud layers, where grains are larger in size, become increasingly affected. The distribution function resulting from the Coulomb explosion of charged dust cloud particles in a plasma can be given by
\begin{equation}
g(a,z)=\left\{\begin{array}{ll}\displaystyle n^{\rm seed}_{\rm d}\delta(a-a_{\rm seed})
&
 \mbox{$a\in[a_{\rm seed},a^{\rm Coul}_{\rm crit})$ } \\
f(a,z) & \mbox{$a\in[a_{\rm crit}^{\rm Coul},a_{\rm max}]$ }\end{array}
\right. 
\end{equation}
where,
\begin{eqnarray}
n^{\rm seed}_{\rm d}&=&n_{\rm d0}\left( \frac{\langle a\rangle_{0}}{a_{\rm seed}}\right)^{3}, \\
n_{\rm d0}&=&\int^{a^{\rm Coul}_{\rm crit}}_{0}f(a,z)~\textnormal{d}a, \\
\langle a\rangle_{0}&=&\frac{1}{n_{d0}}\int^{a^{\rm Coul}_{\rm crit}}_{0}af(a,z)~\textnormal{d}a.
\end{eqnarray}
Note that if the Coulomb explosion process manages to destroy all the large dust particles, all that will remain is a dust cloud composed of nanometre seed particles.

\subsection{Observational consequences of electrostatic disruption \label{sec_obs}}

\begin{figure}
    \centering
        \centering
        \resizebox{\hsize}{!}{\includegraphics{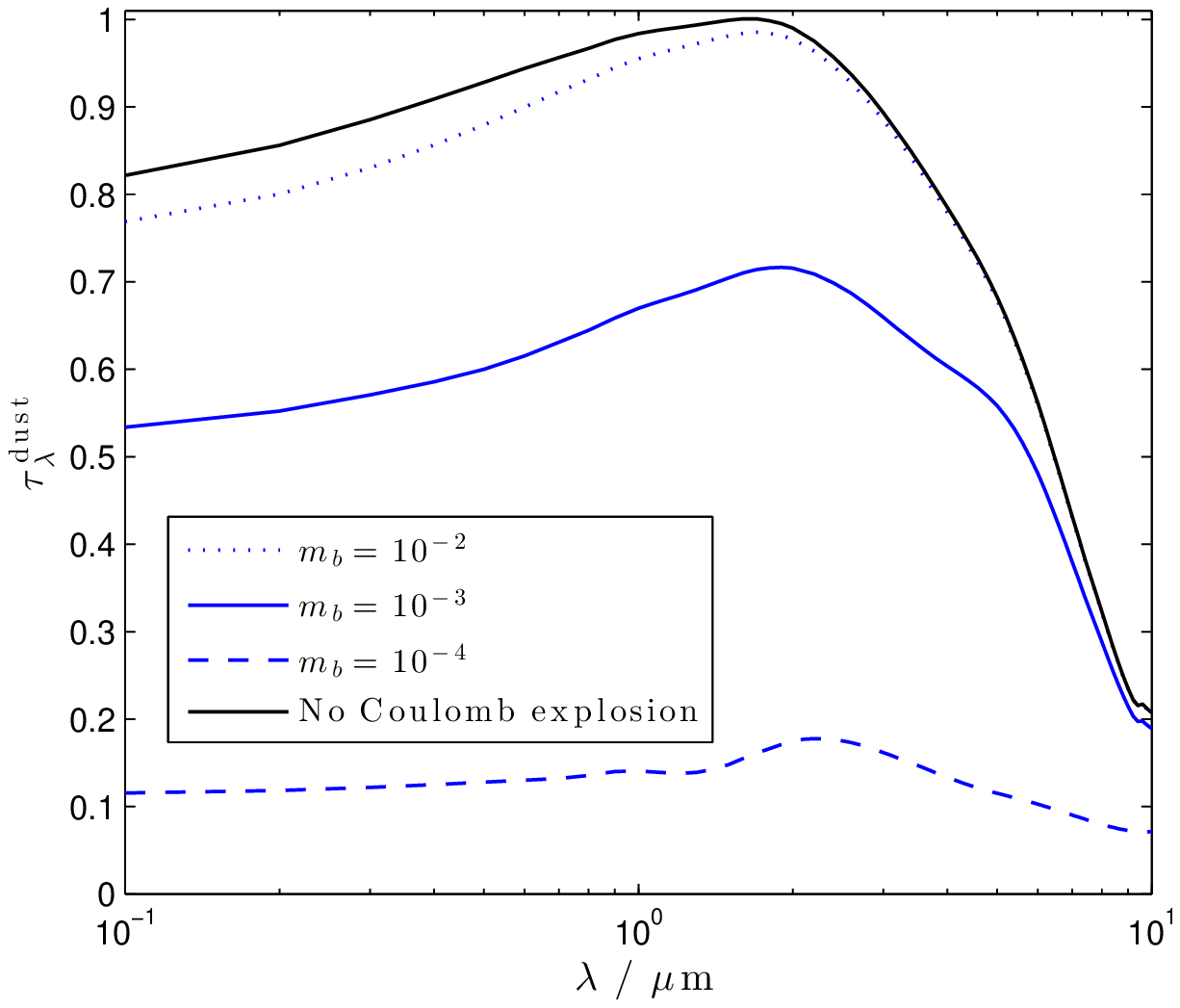}}
    \hfill
        \centering
        \resizebox{\hsize}{!}{\includegraphics{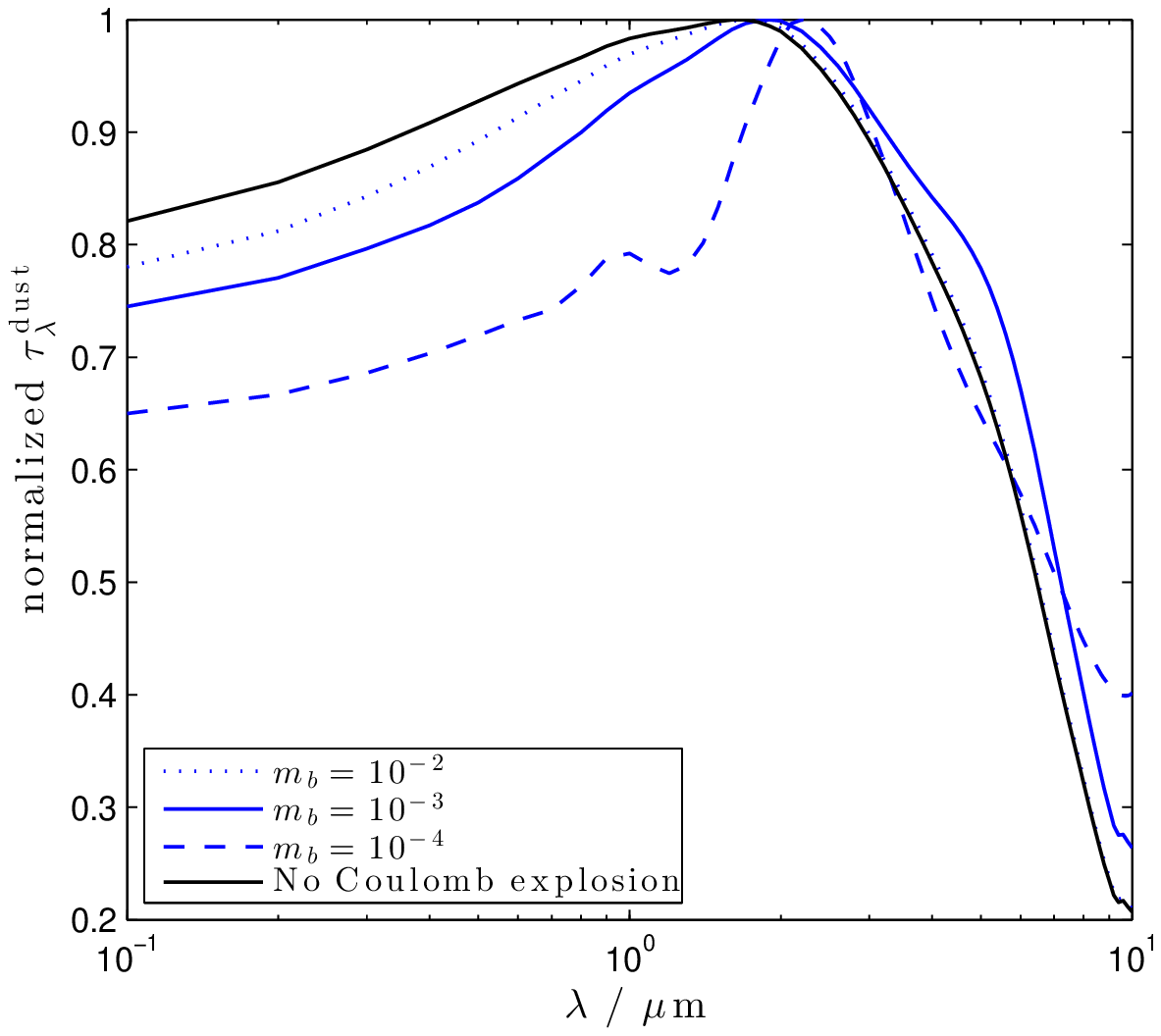}}
    \caption{Dust optical depth $\tau^{\rm dust}_{\lambda}$ for an atmospheric plasma region where the Coulomb explosion of dust occurs. The top figure shows the absolute dust optical depth and the bottom figure shows the dust optical depth normalized by its maximum value so that a comparison of their profiles can be made.}
    \label{opt_depth}
\end{figure}

The resulting optical depth, $\tau^{\rm dust}_{\lambda}$, due to the particle size distribution of the dust can be calculated by evaluating the following double integral,
\begin{equation}
\tau^{\rm dust}_{\lambda}=\int^{z}_{0}  \int_{0}^{a_{\rm max}} g(a,z')~\pi a^{2}Q_{\rm ext}\textnormal{d}a~ \textnormal{d}z',
\end{equation}
where $Q_{\rm ext}=Q_{\rm ext}(a,\lambda,V_{s})$ is the extinction efficiency of the dust, which is a function of the dust particle size, wavelength and volume fraction $V_{s}$. The optical depth is of interest since it determines the intensity (and therefore the flux density for a given solid angle) that is observed. The contrast in optical depth between different atmospheric regions will be seen as a change in the observed flux density. Figure~\ref{opt_depth} shows the resulting dust optical depth as a function of wavelength ($0.1-10~\mu$m) for a region of the atmosphere where the Coulomb explosion of dust particles occurs. To exemplify the Coulomb explosion model an atmospheric plasma region stretching from the bottom of the cloud to the top is considered, using $a_{\rm crit}^{\rm Coul}$ for $T_{e}=10$~eV (Bohm) to calculate $g(a,z)$ and $\tau^{\rm dust}_{\lambda}$ for a lognormal distribution function. To calculate the extinction efficiency, a bespoke Mie scattering numerical routine was utilized using the Bruggeman approximation~\citep{bruggeman1935} to calculate the effective optical constants of the dust grains using values from the references listed in Table~\ref{tbl-3}.  The optical depth is calculated using a Simpson's numerical integration routine over the entire dust particle size distribution function $g(a,z)$ and over the entire cloud extent in $z$. 

As $m_{\rm b}$ decreases in size and the effect of Coulomb explosions increase, more and more dust grains are broken up, leaving a thinner cloud structure with a lower optical depth with respect to adjacent cloud regions where the process is absent. As a result, the contrasting electromagnetic properties of the neighbouring regions would manifest itself observationally as patchy cloud coverage. For comparison, the optical depth for an atmospheric region unaffected by electrostatic disruption is shown (black solid line). 

For clouds with a distribution of particle sizes, dust grains with a length scale comparable to (or harmonics of) the wavelength of the interacting electromagnetic waves resonate most effectively with the radiation. The general profile of the dust optical depth $\tau^{\rm dust}_{\lambda}$ as a function of wavelength $\lambda$ is determined by the superposition of resonant interactions between the electromagnetic radiation and the dust particles. We are interested in photometric variability in the infrared wavelength range corresponding to dust particles with diameters in the micrometer range.  As a result, the optical depth peaks at a wavelength, where the dust cloud is optically thicker, given by $\lambda_{\rm peak}\approx2\langle a\rangle_{*}$ where, 
\begin{equation}
\langle a\rangle_{*}=\frac{\int_{a_{\rm crit}^{\rm Coul}}^{a_{\rm max}}ag(a,z)\textnormal{d}a}{\int_{a_{\rm crit}^{\rm Coul}}^{a_{\rm max}}g(a,z)\textnormal{d}a}.
\end{equation}
Note that in the wavelength range of interest ($\lambda=0.1-10~\mu$m) the effect of the population of seed particles ($a_{\rm seed}\approx\mathcal{O}(1$~nm)) is negligible relative to the effect of the micrometer-sized population, since Mie scattering ($\lambda^{-1}$ for $a\approx\lambda$) dominates over Rayleigh scattering ($\lambda^{-4}$ for $a\ll\lambda$).  As $m_{\rm b}$ decreases, the critical radius $a_{\rm crit}^{\rm Coul}$ increases, the extent of the truncation of the particle size distribution increases and the resulting mean particle size, $\langle a\rangle_{*}$, increases, shifting the peak in the optical depth to higher wavelengths.  Further to this, since the dust particle population is depleted, the cloud becomes optically thinner.  In the case $m_{\rm b}=10^{-4}$ the truncation of the particle size distribution is more severe with $a^{\rm Coul}_{\rm crit}\approx7\times10^{-5}$~cm to $a_{\rm max}$, completely reducing the upper cloud layers to only an ensemble of nanometre sized dust particles. In addition to the population of seed particles at $a_{\rm seed}$, the particle size distribution in lower cloud layers is restricted to a narrow distribution of larger dust grains.  As a result of this narrow distribution, there is insufficient overlap in the resonant features at wavelengths corresponding to harmonics of the dust diameter, to guarantee a continuous optical depth profile;  hence, the dip in the optical depth at $\lambda\approx 1.2~\mu$m. This is in contrast to the cases $m_{\rm b}=10^{-3}-1$, where the spread in particle sizes is sufficiently broad that the dust-radiation resonant features form a continuous optical depth profile.  In the extreme case where all dust particles fall below the critical threshold for stability, only a cloud composed of nanometre-sized seed dust particles will exist making the cloud optically thin in the wavelength range presented here ($0.1-10~\mu$m) with a characteristic Rayleigh scattering intensity dependence.  These features present useful diagnostics for distinguishing the Coulomb explosion process from other inhomogeneous cloud coverage mechanisms.

\section{Discussion \label{discussion}}
The aim of this paper was to investigate the electrostatic disruption of cloud dust particles in substellar atmospheres and to assess the contribution to inhomogeneous cloud coverage and the resulting spectroscopic variability.  Dust cloud particles immersed in a plasma region, with a radius below a critical radius $a<a_{\rm crit}^{\rm Coul}$ are electrostatically disrupted and broken apart. This results in a bimodal particle size distribution composed of stable nanoscale seed particles and particles with $a\geq a_{\rm crit}^{\rm Coul}$, with the intervening range of particle sizes defining a region of instability devoid of dust.  For an atmospheric plasma region with an electron temperature of $T_{e}=10$~eV, the critical grain radius varies from $a^{\rm Coul}_{\rm crit}\approx10^{-7}$~cm to $10^{-4}$~cm depending on the tensile strength of the dust particle, characterized by $m_{\rm b}=1$ to $10^{-4}$ respectively. Higher critical radii are attainable for higher electron temperatures.  As $m_{\rm b}$ decreases, the dust population is depleted and the clouds become optically thin. Furthermore, as the critical radius $a_{\rm crit}^{\rm Coul}$ increases, the extent of the truncation of the particle size distribution increases and the resulting mean particle size increases, shifting the wavelength-particle size resonant peak in the optical depth to higher IR wavelengths. Depending on the distribution of the dusty plasma regions in the substellar atmosphere, electrostatic disruption of dust clouds could yield regions of contrasting cloud and optical properties, which gives a source of inhomogeneous cloud coverage that could be associated with the observed spectroscopic variability.  If the entirety of the atmosphere is ionized then all the atmospheric dust is charged and susceptible to electrostatic disruption.  In this case inhomogeneous cloud coverage would occur due to non-uniformity in the electron temperature across the atmosphere since $a^{\rm Coul}_{\rm crit}$ is dependent upon the electron temperature. 

The electrostatic disruption of dust grains in substellar atmospheres has a significant impact on the subsequent chemistry that occurs. Consider a spherical dust grain of radius $a$ and corresponding surface area. If such a grain is electrostatically disrupted it will break-up into numerous smaller fragments each with a smaller surface area than the initial, parent dust grain. However, the resultant combined surface area of the fragment ensemble will be greater than that of the initial dust grain. As a result, reagents and radicals have a greater surface area available to them, promoting surface catalysis. The increased collective surface area from fragmentation is in analogy with micronized coal-fired power generation: lumps of coal are broken-up into micrometer sized particulates (via a collisional process) the collective surface area of the micronized coal is greater than that of the parent lump. Therefore, surface catalysis is enhanced and the efficiency of combustion increased resulting in a greater energy yield in comparison to standard coal-fired stations. Although the specific chemical reactions in the micronized coal context are different, the underlying physical principle is the same: an increased surface area enhances the chemical reactivity.  The increased reaction surface area may cause a stronger depletion of the gas-phase species that participate in grain surface chemistry (see Table 1, \cite{helling2008c}). Hence, localized time-dependent gas-phase depletion may well be caused by a sudden population boost of small grains induced by electrostatic disruption of larger grains. 

The electrostatic disruption of dust grains may also be applicable in other astrophysical, dusty plasma environments, such as supernova remnants. Supernova remnants are associated with the creation of a series of shocks that propagate through the dusty ejecta affecting dust formation and destruction. Investigations have suggested that grains with a length scale $\lesssim 0.05~\mu$m are destroyed by sputtering in the post-shock flow while those  $\approx0.05-0.2~\mu$m are trapped in the shock~\citep{nozawa2007}.  The fraction of dust that survives the passage of the shock ranges between 2-20\%~\citep{bianchi2007}, which may be further affected by additional processes such as the electrostatic disruption of dust. Furthermore, observed silicate features in protoplanetary disks are believed to be consistent with the existence of a population of micrometer sized grains and the absence of sub-micrometer sized grains~\citep{williams2011}. To produce such a particle size distribution,~\cite{olofsson2009} suggest that either dust growth processes dominate fragmentation at these length scales or the sub-micrometer dust grains are removed from the upper layers of the disk via winds or radiation pressure.  The electrostatic disruption of dust particles could be an additional mechanism aiding in the removal of the sub-micrometer particles from the upper layers of the disk. 

\begin{table*}
\caption{Tensile strengths of selected materials.\label{tbl-2}}
\centering
\begin{tabular}{l c c}
\hline \hline
Material & Tensile Strength (MPa) &  \\
\hline
Fe &10 &\cite{khol1995}\\
TiO$_{2}$ & 36.41 & \cite{qiu2006}\\
Al$_{2}$O$_{3}$ &241.4&  \cite{shackelford2001}\\
Mg$_{2}$SiO$_{4}$ &9.5 & \cite{petrovic2001}\\
 SiO$_{2}$& 48& \cite{lynch1983}\\
MgSiO$_{3}$ & 9.6 & \cite{popova2011}\\
MgO &96 & \cite{stokes1963,shackelford2001}\\
\hline
\end{tabular}
\end{table*}
\begin{table*}
\caption{Reference list for the optical constants of the materials used to calculate the extinction efficiency, $Q_{\rm ext}$ .\label{tbl-3}}
\centering
\begin{tabular}{l l}
\hline \hline
Material & Reference   \\
\hline
Fe & \citet{ordal1988} \\
TiO$_{2}$ & \citet{posch2003,zeidler2011} \\
Al$_{2}$O$_{3}$ & \citet{palik1991,zeidler2013} \\
Mg$_{2}$SiO$_{4}$ & \citet{jager2003} \\
 SiO$_{2}$&  \citet{palik1985,zeidler2013} \\
MgSiO$_{3}$ & \citet{jager2003}  \\
MgO & \citet{palik1991} \\
\hline
\end{tabular}
\end{table*}
\begin{acknowledgements}
The authors are grateful to the anonymous referee for constructive comments and suggestions that have improved this paper. ChH and CRS are grateful for the financial support of the European Community under the FP7 by an ERC starting grant number 257431.  DAD is grateful for funding from the UK Science and Technology Funding Council via grant number ST/I001808/1.  It is a pleasure to acknowledge stimulating discussions with Drs H. E. Potts and P. B. Rimmer.  Further to this, the authors would like to thank D. Juncher for discussions regarding dust opacity. The authors would like to acknowledge our local IT support.
\end{acknowledgements}
\end{document}